\documentclass[sn-mathphys-num]{sn-jnl}%

\usepackage{multirow}%
\usepackage{amsmath,amssymb,amsfonts}%
\usepackage{amsthm}%
\usepackage{mathrsfs}%
\usepackage[title]{appendix}%
\usepackage{xcolor}%
\usepackage{textcomp}%
\usepackage{manyfoot}%
\usepackage{booktabs}%
\usepackage{algpseudocode}%
\usepackage{listings}%
\usepackage{subcaption}
\usepackage{comment}
\usepackage{array}
\usepackage{graphicx}
\usepackage{booktabs}
\usepackage{xcolor}
\usepackage{colortbl}
\newcolumntype{P}[1]{>{\centering\arraybackslash}p{#1}}

\usepackage{graphicx}
\usepackage{float}

\begin{document}
\let\WriteBookmarks\relax
\def\floatpagepagefraction{1}
\def\textpagefraction{.001}

\title {Transition of car-based human-mobility in the pandemic era: Data insight from a cross-border region in Europe}                      

\author*[1]{\fnm{Sujit Kumar} \sur{Sikder}}\email{s.sikder@ioer.de}
\author[2]{\fnm{Jyotirmaya} \sur{Ijaradar}}\email{jyotirmaya.ijaradar@tu-dresden.de}
\author[3]{\fnm{Hao} \sur{Li}}\email{hao.li@nus.edu.sg}
\author[4]{\fnm{Hichem} \sur{Omrani}}\email{hichem.omrani@liser.lu}

\affil[1]{\orgname{Leibniz Institute of Ecological Urban and Regional Development (IOER)}, \city{Dresden}, \country{Germany}}

\affil[2] {\orgname{TU Dresden}, \city{Dresden}, \country{Germany}}

\affil[3]{\orgname{National University of Singapore}, \city{Singapore}, \country{Singapore}}

\affil[4]{\orgname{Luxemburg Institute of Socio-Economic Research (LISER)}, \city{Esch-sur-Alzette/Belval}, \country{Luxembourg}}

\abstract{
Many transport authorities are collecting and publishing almost real-time road traffic data to meet the growing trend of massive open data, a vital resource for foresight decision support systems considering deep data insights. We explored the spatio-temporal transitions in the cross-country road traffic volumes in the context of modelling behavioural transitions in car-based human mobility. This study reports on individual car-based daily travel behaviour detected, before (2018) and during the COVID pandemic (2020), between Germany and neighbouring countries. In the case of Luxembourg, the Bridges and Roads Authority has installed a large digital traffic observatory infrastructure through the adoption of sensor-based IoT technologies, like other European member states. Since 2016, they have provided high-performance data processing and published open data on the country's road traffic. The dataset contains an hourly traffic count for different vehicle types, daily for representative observation points, followed by a major road network. The original dataset contains significant missing entries, so comprehensive data harmonization was performed. We observed the decrease in traffic volumes during pandemic factors (e.g. lockdowns and remote work) period by following global trend of reduced personal mobility. The understanding the dynamic adaptive travel behaviours provide a potential opportunity to generate the actionable insight including temporal and spatial implications. This study demonstrates that the national open traffic data products can have adoption potential to address cross-border insights. In relevance to the net-zero carbon transition, further study should shed light on the interpolation and downscaling approaches at the comprehensive road-network level for identifying pollution hot spots, causal link to functional landuse patterns and calculation of spatial influence area.   
}


\maketitle

\section{Introduction}

Transition in human mobility behaviour is a critical topic within the growing convergence research initiatives on sustainable spatial development, climate protection, and the transformation of society \citep{Sachs2019, Sekadakis2023}. These require innovative approaches to understanding and managing human mobility, particularly because car-based mobility is a major contributor to greenhouse gas emissions \citep{SHAHEEN2007}. The advancements in digital data-sharing culture, e.g. open data infrastructures, are creating further opportunities for evidence-informed policymaking for robust knowledge generation \citep{Zuiderwijk2019}. For instance, the availability of high-frequency open traffic data is unlocking the potential to develop analytics with a replicable pipeline that provides deeper insights and efficiency in policy-making. 

The spatial-temporal change analysis of human mobility behaviour is complex and multifaceted, and it is shaped by a wide range of factors, including demographic, cultural, and economic variables, as well as technological developments and policy decisions. The COVID-19 pandemic has drastically altered our mobility behaviour and model choices. As a result of public health measures mandating physical distancing and avoiding crowded areas, mobility patterns have shifted significantly over the past year. Some research has even claimed that the new mobility transition has also been impacted by the so-called “Urban exodus” - where large numbers of the city population are leaving the city area and starting to live in rural areas. \cite{dePalma2022} reviewed and discussed the primary influences on short-term, medium-term, and long-term mobility decisions, including route, departure time, mode, teleshopping, teleworking, car ownership, work location, choice of job, and residential location. Once centralised in offices and shopping malls, activities have shifted to home-based remote work and online shopping \citep{Caselli2022, Clark2021, huang2022staying}. These changes have necessitated the implementation of hygiene regulations to maintain safe distances and avoid mass gatherings, resulting in a significant shift from public transit to individual car-based human mobility. Studies are already reporting transitions (pre- and during COVID-19) in mobility sectors using diverse data sources, including active and passive data sources \citep{Macharis2021, Schmidt2021}. For example, the dynamics of Points of Interest (POI) attractiveness have been analysed using Google Mobility data  \citep{Psyllidis2022}, and social media data has been employed to gauge mobility changes \citep{Zachlod2022, AcostaSequeda2024} and so on. Many of these studies have limited focus on the cross-border spatial dimensions using open data \citep{Connelly2023}, a transparent data harmonisation approach, addressing biases in data quality \citep{FernandezArdevol2022}.

This study aims to present an exploratory approach to analysing spatio-temporal transitions in road traffic volumes as a proxy for modelling behavioural changes in car-based human mobility, including cross-country analysis. This study contributes to developing an analytical framework for car-based human mobility behaviours, demonstrated through an empirical case study using Luxembourg's road traffic count data. These analytics are instrumental in supporting evidence-informed decisions for land use policy measures, emission reduction strategies, air pollution risk management, and regional development policies, even though traffic count data is only available as open data from Luxembourg.  This study demonstrates that the national open traffic data products can have adoption potential to address cross-border (e.g. Germany does not have Open traffic data, but Luxembourg has open data)  policy and environmental challenges associated with car-based human mobility in the pandemic era and beyond. The results and methodologies developed in this study are flexible to adopt and replicable for other cases, provided that comparable traffic data is available.  Our proposed workflow has followed the requirements of open science principles by adopting open data, delivering an open-source Python package and reproducible notebooks of analytical outputs. 

\section{Related Works}

One of the most notable changes in car-based travel behaviour was the reduction in the number of trips taken overall. While the COVID-19 pandemic has led to some changes in car-based travel behaviour, the long-term impacts on travel patterns and preferences are still uncertain \cite{Andani2024, Lodi2024}. \cite{Shamshiripour2020} observed considerable changes in people's mobility styles and routine travel habits based on a survey done in the Chicago metropolitan region. The survey includes a series of questions regarding the travel activities, habits, and impressions of individuals before and during the pandemic, as well as their future expectations. Similar findings were discovered in many other countries around the world, including Pakistan \citep{Abdullah2020}, Turkey \citep{Shakibaei2021}, Switzerland \citep{Molloy2021}, Bangladesh \cite{Mahmud2024},  Canada \cite{Haseeb2024} and the Netherlands \cite{deHaas2020}. 

\cite{Eisenmann2021} found that public transportation has been particularly affected in Germany, as people avoid buses and subways due to health and safety concerns. This has led to an increase in the use of personal vehicles, with car ownership becoming a necessity for many. The COVID-19 crisis became a catalyst towards sustainable mobility transition \citep{Macharis2021, Schmidt2021}. The pandemic prompted a shift away from public transportation, such as buses and trains, due to concerns about virus transmission. \cite{Ecke2022} agree with the declining popularity of public transport and also reported that commuters in Germany who commuted during the pandemic did not notably change their commuting behaviour in terms of commuting time or mode. However, Cai et al. (2024) found that public transport use remained unchanged among low-income groups, while driving alone increased among high-income workers.  \cite{Macharis2021} explored the scope of the possible sustainable mobility transition by studying the Covid-19 crisis using the 7A's framework for sustainable mobility, which stands for awareness, avoidance, action, and shift as well as anticipation of new technologies and actor participation, as well as acceleration and behaviour adaptation. \cite{Schmidt2021} conducted an online survey method in a sample representative of the population to investigate the effects of the pandemic-related restrictions on daily life, travel mode choices, as well as their future mobility desires. \cite{Kellermann2022} discussed the longitudinal analysis of changing urban mobility behaviours before and during the COVID-19 pandemic and concluded that during the pandemic, people travelled shorter distances, travelled less frequently, and switched to active modes.

Exploratory Data Analysis (EDA) is a well-established approach of summarising and visualising data to gain insights and comprehension about the underlying structure of data. Identifying patterns, relationships, and potential anomalies within the data is a crucial step in the data analysis process \citep{Behrens1997, Ghosh2018, Hullman2021}. In the context of traffic count data, EDA may entail analysing the distribution of vehicle counts, identifying trends over time, and investigating the relationships between traffic volume and other factors such as weather, road conditions, and travel time. EDA results can be used to improve traffic modelling, support data-driven decision-making, and inform initiatives aimed at reducing congestion and improving mobility \citep{Chaloli2019, Han2006, Soorya2020, Taboada2020}. \cite{Makaba2020} introduced a comprehensive approach to analysing a real-world dataset of road traffic accidents by utilising graphical representations and implementing dimension reduction techniques. The study carried out both Principal Component Analysis (PCA) and Linear Discriminant Analysis (LDA) on such a dataset. The results are measured by performance metrics, and it has offered a thorough understanding of the patterns of road traffic accidents. \cite{Hungness2022} assessed the feasibility of forecasting and evaluating the impact of vehicle miles travelled (VMT) in Dane County (Wisconsin) using existing travel demand models. Their research highlights that a thorough exploratory spatial data analysis of the collected data reveals significant spatial correlations and interactions that traditional methods fail to capture. This information can be used by planners to target areas where VMT remediation efforts for sustainable transportation networks and environments can be most efficiently implemented. 

\begin{table}[h!]
\centering
\caption{Examples of studies that used traffic data}
\begin{tabular}{|p{2cm}|p{3.5cm}|p{1.5cm}|c|c|p{2cm}|p{1.8cm}|}
\hline
\textbf{Study} & \textbf{Aim} & \textbf{Data} & \textbf{Spatial} & \textbf{Temporal} & \textbf{Application scale} & \textbf{Empirical study} \\
\hline
\cite{Hamza2023} & Road traffic jam studied by bridging between network factors and motorist behaviour & Traffic jam incidents & No & Yes & City & Kampala, Uganda \\
\hline
\cite{Jin2024} & Assess the impact of Shanghai’s 2022 COVID-19 lockdown on traffic and related socio-economic indicators & Traffic volumes, metro ridership & Yes & Yes & City & Shanghai, China \\
\hline
\cite{Baccoli2022} & High time resolution traffic noise predictions using nonlinear ANN & Traffic count at road lane & Yes & Yes & City & Italy \\
\hline
\cite{Makaba2020} & Proposed and validated EDA framework for road traffic accidents - graphical representations, dimensionality reduction methods & Government Road traffic safety data & No & Yes & Regional (Province) & South Africa \\
\hline
\cite{Soorya2020} & EDA using ML reported K-KNN non-parametric regression performed better than RF, MCS, SARIMA & Traffic count at road toll-way & No & Yes & Regional & India \\
\hline
\cite{Chaloli2019} & EDA in 4 steps - choice of the right dataset, cleaning, exploratory analysis, drawing inferences for deriving insights on the work culture, migration patterns, etc. of the locality & Traffic count city data & Yes & Yes & City & New York, US \\
\hline
\cite{Shamo2015} & Application of kriging and variogram algorithms analysis and predictive modelling of road traffic & Annual average daily traffic (AADT) & Yes & Yes & Regional (State) & Washington, US \\
\hline
\cite{Sun2023} & Investigation of whether public sentiment, derived from Twitter data, can be used to better explain and predict changes in urban mobility during the COVID-19 pandemic & Twitter data and Google COVID-19 Community Mobility Reports & Yes & Yes & City & Kyoto, Japan\\
\hline
This study & Exploratory approach to analyze the spatiotemporal changes in the road traffic volumes as a proxy for modelling behavioural changes in car-based human mobility including cross-country analysis & IoT-based traffic count observatory & Yes & Yes & Country, cross-border & Luxembourg and Germany \\
\hline
\end{tabular}
\label{tab:example_studies}
\end{table}

Table ~\ref{tab:example_studies} shows some example studies that adopted the exploratory data analysis (EDA) approach for road traffic data-related studies around the world. This study contributes to the existing knowledge by demonstrating an exploratory approach to the analysis of spatiotemporal changes in road traffic volumes as a proxy for modelling behavioural changes in car-based human mobility, including cross-country analysis.

\section{Analytical framework for sensing road traffic volumes at the cross-border region}

An emerging form of data, including open big data, has the potential to contribute to digitally innovative solutions for complex societal problems, including environmental justice. Big data is defined mainly by size among other characteristics, normally a very large and complex dataset \citep{Brunswicker2015, Kitchin2016}. While open data is defined by its use, a public and easily accessible dataset can be used for a new trend or pattern analysis. Common sources for big open data on mobility topics are:  (i) User-generated data are any form of information like feelings, opinions, comments, reviews, and images generated by an individual user \citep{Saura2021}. In the mobility sector, the traditional user-generated data are mainly user survey data, but now there are also large amounts of user data collected from transit apps, smart cards and so on. (ii) Transaction-generated data is obtained from commercial and nonprofit transactions involving persons in an expanding number of computerised day-to-day activities \citep{Krishnan2013}. In mobility research, this type of data is collected from transport subscriptions, toll taxes, and petrol pump transactions. (iii) Sensor-generated data are generated by different sensors without human intervention; often networked and collected near real-time \citep{Teh2020}. This type of data source is the largest contributor to open mobility data by using traffic cameras and roadside sensors.
Road traffic data or traffic count data can be collected in both automated and human-centric ways. The modes can be categorised into 4 types: (i) using road sensors, (ii) installing traffic cameras, (iii) adopting probe-based collection technologies, and (iv) administering household travel surveys based on their development and working principles \citep{BITREIPA2014}. We contribute to the current literature by considering the automated mode of data source that collects high-frequency road traffic data from IoT sensor network observatory in combination with multiple sensors, cameras and advanced technologies. Our analytical approach considers the following dimensions: 
\begin{enumerate}
    \item \textbf{Data quality dimension:} This has to deal with various data visualisation and statistical techniques in understanding data quantity, insightful distribution for day, week, month and year, as well as the locational factors of traffic counting stations.
    \item \textbf{Data harmonization dimension:} The data insights have to be exploited deeper after an initial exploration of data quality dimensions at the entire dataset; the focus on specific areas of interest and prior sub-setting can also be decided.  The major issues have to be addressed, such as missing entries, unequal entry counts, aggregation at temporal resolution – day, weeks, months, type of cars and combined traffic-ways. Secondary data creation should be useful for meaningful analysis and easy computation.
    \item \textbf{Distributional dimension:} The distribution of traffic needs to be understood by exploring emerging trends and hot-spot detection. So, the spatial intensity of the traffic counts, clusters and significant patterns can be observed at the local as well as national scale. 
    \item \textbf{Directional cross-border dimension:} The temporal variation differs in the traffic volume following the directions of traffic flow; however, this can be hard to achieve with traffic count at the inner- country-level (modelling can be approached by including complex functional landuse, comprehensive parameters of road networks and so on) due to data quality and appropriate location of the observatory. However, the cross-border traffic flow trends are near the international borders with other neighbouring countries. Therefore, the major entry point observation stations' data should be considered, which are near motorways during rush hours. This output should act as a proxy for capturing daily commuters for diverse trip purposes – work, shopping, recreation and so on. 
    \item \textbf{Case study dimension:} Our analytical framework is applied to a practical case study in Europe – Luxembourg. The country experiences intensive and frequent commuter traffic from/to neighbouring nations, Germany, France, and Belgium.
\end{enumerate}

\section{Description of the case study, data and workflow}
The Luxembourg Roads and Bridge Corporation published several open datasets related to national mobility and traffic observations. Two major datasets are - PCH (National Road Traffic Count dataset) and CITA (Traffic Events in DATEX II). We considered the PCH data and used the annual traffic count data for 2018 and 2020. The PCH annual traffic count dataset is an annually published public dataset containing hourly traffic counts for cars, trucks, or both in each direction and different routes throughout Luxembourg. It is sensor-generated data and has 36 features: position ID, route, vehicle type, hourly traffic count, and direction. The dataset is downloadable from a permanent link provided by Luxembourg Roads and Bridge Corporation in XLSX format, and the data has 457,098 entries in 2018 and 504,886 entries in 2020 Table ~\ref{tab:dataset}. 

\begin{table}[h!]
\centering
\caption{Description of the dataset}
\begin{tabular}{|l|c|c|}
\hline
\textbf{Characteristics} & \textbf{2018} & \textbf{2020} \\
\hline
Size of the dataset & 457,098 & 504,886 \\
Number of features & 36 & 36 \\
Observation stations & 176 & 178 \\
Routes & 80 & 78 \\
All types of Vehicles (count) & 532,244,275 & 477,034,465 \\
Cars only (count) & 35,454,836 & 33,771,145 \\
\hline
\end{tabular}
\label{tab:dataset}
\end{table}

The raw dataset was processed and analysed in the following major steps described below. 

\begin{enumerate}
\item \textbf{Step 1:} Focuses on exploring the data through various data visualization and statistical techniques to understand its size, accuracy, and overall quantity. To gain insights into the nature of the data, we utilise a calendar chart to analyze the data distribution and identify any missing data for each week of every month throughout the year.
\item \textbf{Step 2:} After reviewing the entire dataset, we applied filters to refine it to our area of interest. To deal with the missing entries and unequal entry counts between the datasets, we created an aggregated dataset by averaging weekday values for one week from each month. We further narrowed this down by applying filters specifically for cars and two-way traffic. To enhance our analysis, we created two new features, morning and evening rush hours, by analysing car volume between 7-10 am and 4-7 pm, respectively. The final result was a harmonised dataset that allowed for robust and valid analytics in a short computation time.
\item \textbf{Step 3:} Analysed the emerging trends and hotspots to find more densely travelled areas during both morning and evening rush hours for the whole country of Luxembourg. This helped to understand the traffic flow trends on a national scale.
\item \textbf{Step 4:} Conducted the cross-border traffic flow analysis to explore more evidence for temporal traffic flow trends near international borders. Luxembourg has three neighbouring countries: Germany, France, and Belgium, and is connected by motorways (A1, A3, A7). The traffic volumes at entry point observation stations near these motorways were analysed during rush hours.
\item \textbf{Step 5:} Analyse the cross-border traffic flows between Germany and Luxembourg. Germany shares about 138 km of border with Luxembourg; therefore, the cross-border observation station was selected between Germany and Luxembourg. For this analysis, we considered one-way traffic to understand the traffic flow pattern during rush hours to and from Germany. We also observed the day-wise hourly traffic flow to understand car-user travel patterns between these two countries. This analysis estimates daily commuters for work-related trips between Germany and Luxembourg.
\end{enumerate}

\begin{figure}
    \centering
    \includegraphics[width=0.85\linewidth]{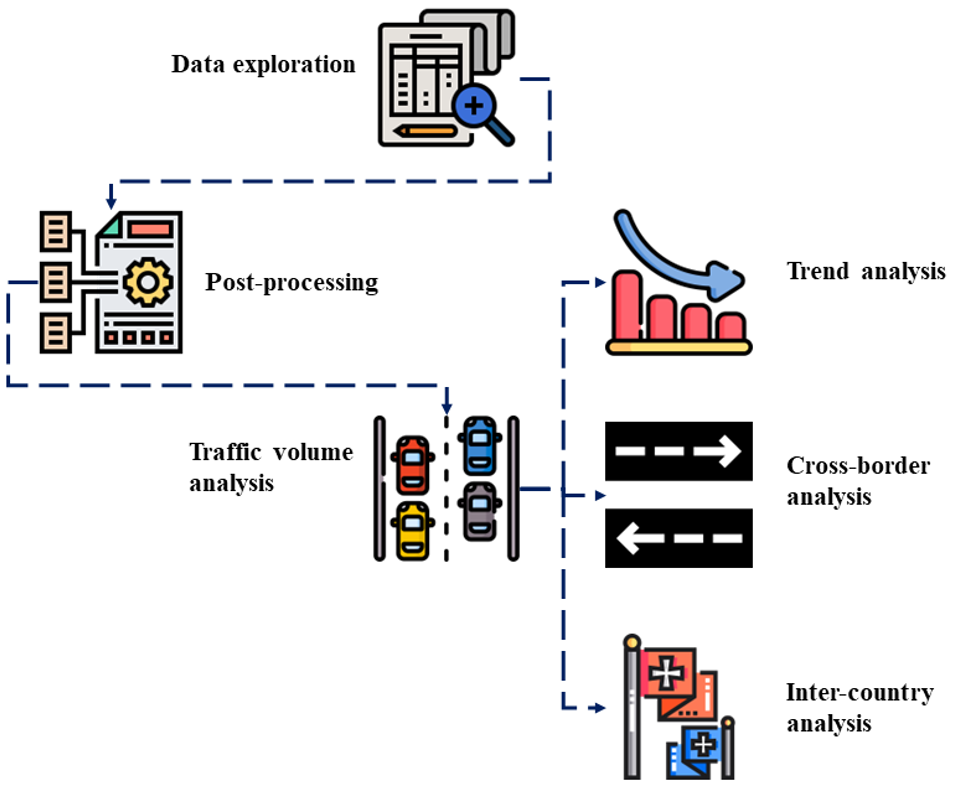}
    \caption{Data processing workflow and analytical steps}
    \label{fig:workflow}
\end{figure}

This study contributes a prototype semi-automated workflow using open-source packages of the Python programming language (see open-source code in GitHub \cite{Ijaradar2024}). Figure ~\ref{fig:workflow} displays the workflow for data processing and analytical steps. The following section presents exploratory findings at national and selected observation station levels, which are located at the entry point to Luxembourg near the cross-country border.

\section{Results of the case study}
\subsection{Data exploration}

The study conducted exploratory analytics using the complete annual dataset of digital road traffic count information from 156 observation stations, which were identified by POSTE\_ID. The focus of this study was on car traffic, even though data on other types of vehicles, such as heavy trucks, are also available from the observation points along the main road network. For the years 2018 and 2020, three different analyses were performed. First, monthly aggregated car traffic counts were calculated, and annual calendar heat maps were created to display the results. Second, the trends in weekly total traffic volumes were analysed for a selected week each month, and differences among weekdays, Saturdays, and Sundays were noted.

\subsubsection{Year-wise traffic volume calendar}

The annual calendar heat maps were used to display car traffic volumes during the morning and evening in rush hours, providing valuable insights into the data quality and aiding in the selection of representative weeks for further analysis. Figure ~\ref{fig:cal-morning-evening} shows some incompleteness of data point due to missing values, or COVID-19-related restrictions. These limitations need further investigation beyond simple data aggregation for an entire month or a year, as they may impact the accuracy of reporting changes in road traffic volume. Therefore, aggregation was performed for selected weeks from each month, with a focus on rush hour periods to ensure maximum completeness of the data (Figure ~\ref{fig:cal-morning-evening-sel} shows the days and weeks considered for aggregation).

\subsubsection{Monthly Traffic volume}  

Figure ~\ref{fig:traffic-volume} presents the monthly total traffic volume for the years 2018 and 2020. In 2018, there was a drop in traffic volume during the morning rush hours from the end of July to August. The highest traffic volume was seen in July and September, with a significant decrease in traffic volume observed in September. This trend was also observed in 2020. The summer holidays, which typically include school closures and a suspension of construction activities for four weeks in August, could be one of the reasons for the decrease in traffic volume.
\begin{figure}[h!]
    \centering
    \includegraphics[width=0.85\linewidth]{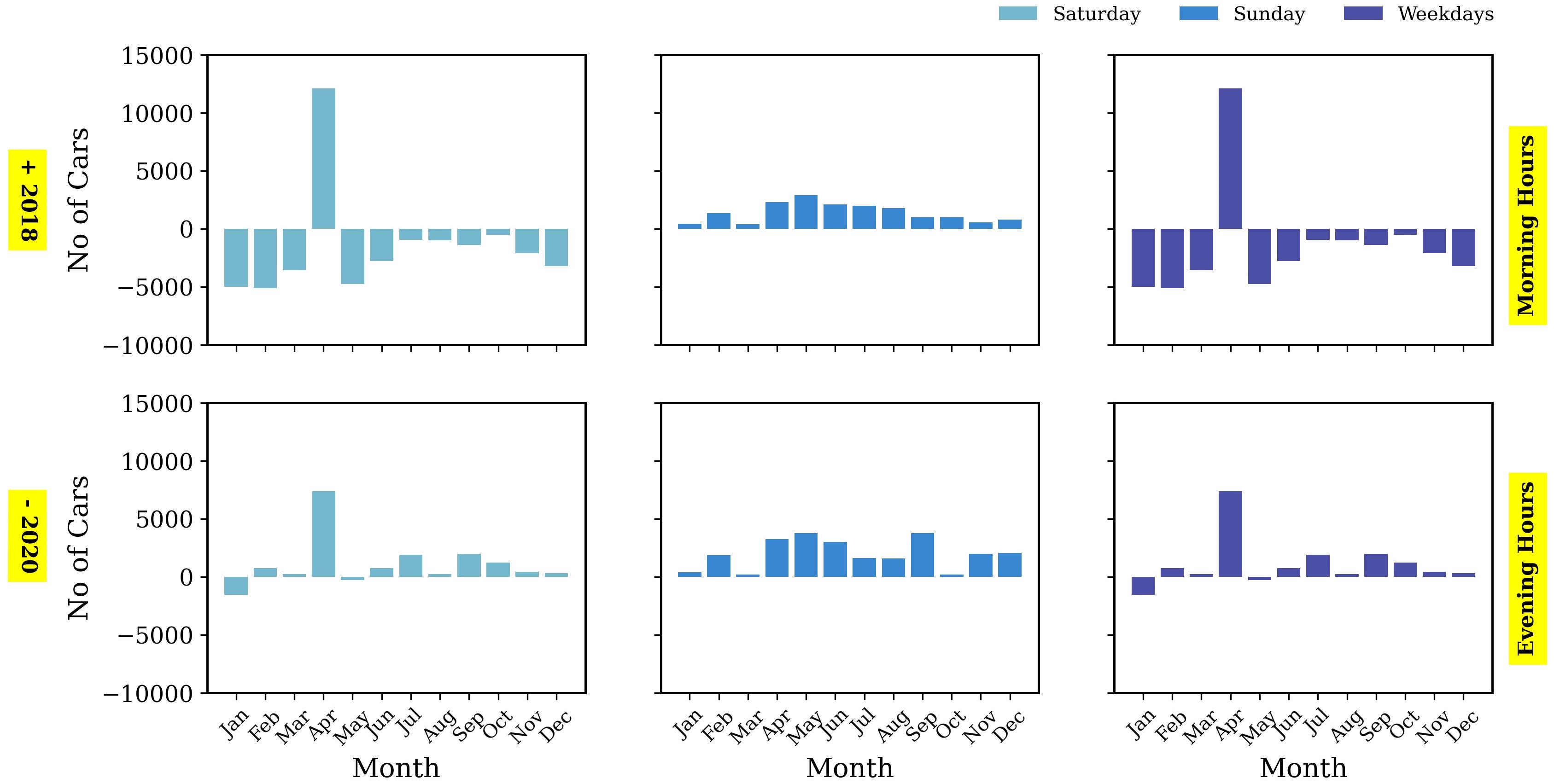}
    \caption{Comparative changes in car-traffic volumes between 2018 and 2020}
    \label{fig:chnages-traffic-volume-18-20}
\end{figure}
 
In 2020, car traffic volume during weekday morning hours hit its lowest point due to COVID-19 restrictions. However, from May to June, the trend started to increase again as COVID regulations were relaxed and summer vacations resumed. The increase in traffic volume continued from June to July. On Sundays, there was an overall decrease in traffic volume during morning hours. A similar trend was observed on both Saturdays and Sundays, although the decrease was more pronounced on Sundays. This could be attributed to shopping activities for daily necessities, as people tend to shop more on weekends. On weekdays, individuals may have been occupied with work-from-home duties, leaving less time for shopping. Figure ~\ref{fig:chnages-traffic-volume-18-20} shows the changing trend of car traffic volume for a firm comparison between 2018 and 2020.

\subsubsection{Hot spot of spatiotemporal variability by observation stations}
\begin{figure}[h!]
    \centering
    \includegraphics[width=0.99\linewidth]{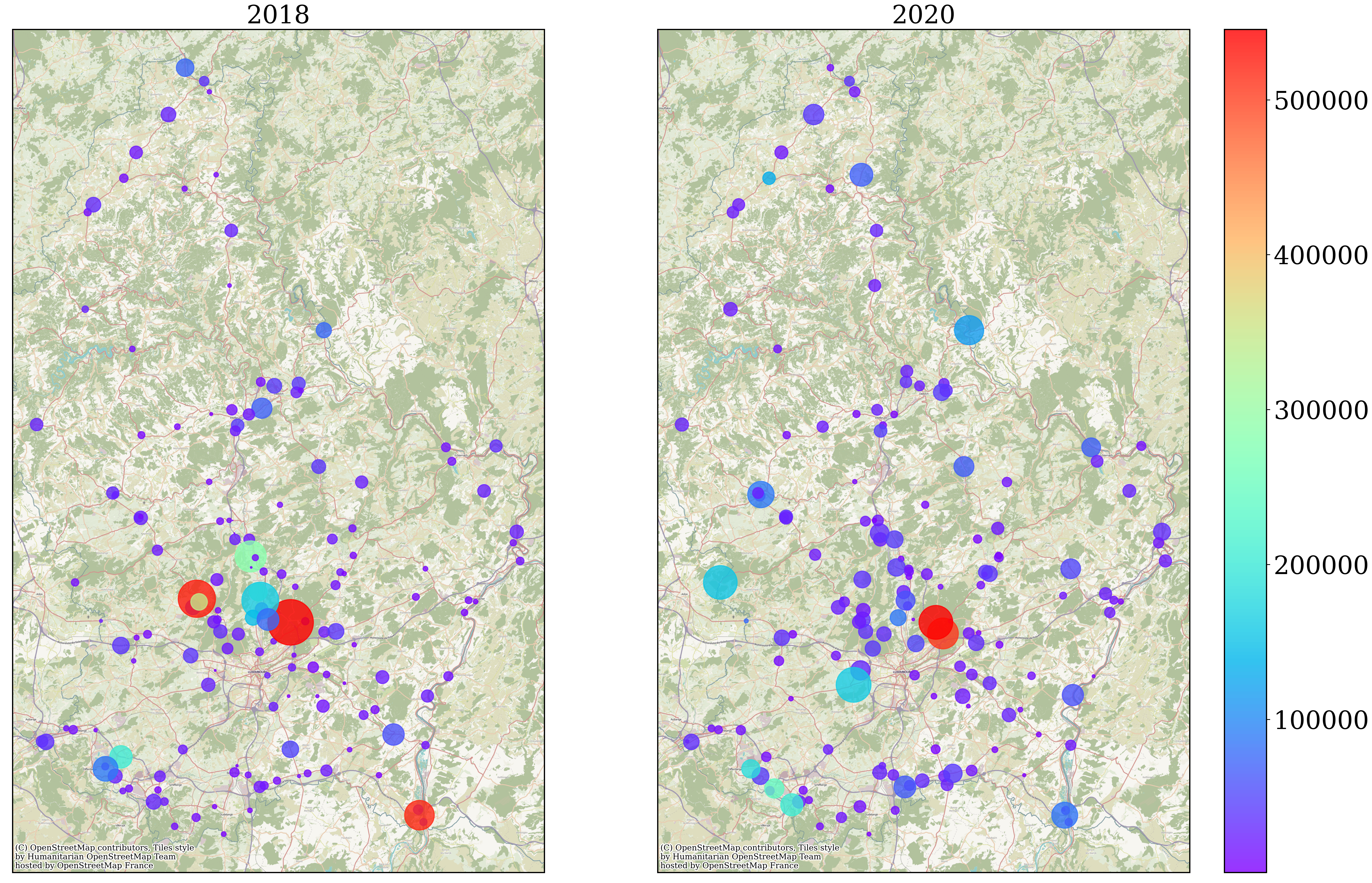}
    \caption{Spatio-temporal variability in car-traffic volumes in the first week of January. Legends notes: Size = difference between morning to evening, colour = total car traffic}
    \label{fig:chnages-mor-eve}
\end{figure}

Figure ~\ref{fig:chnages-mor-eve} shows the spatial variability of car traffic volume and its changes between 2018 and 2020. The figure uses colours to represent the total car traffic, while the size of the bubbles indicates the difference in car traffic volume between morning and evening at a counting location. There are many changes in the car traffic volume between 2018 and 2020. The hotspot of car-traffic changes may be highly linked to the functional land-use structure. This means that areas with certain types of land use, such as residential or commercial, may have different traffic patterns. For example, the highest car traffic volume and changes can be found at the observation stations within or near the city of Luxembourg.

\subsection{Traffic volume in Luxembourg and its neighbouring countries}

Luxembourg is a landlocked country in Western Europe that shares borders with Belgium to the west and north, Germany to the east, and France to the south. Due to its location, many people travel to Luxembourg by car from neighbouring countries. The country has an extensive road network and is well-connected to its neighbours through highways and motorways. On February 1, 2020, Luxembourg introduced a new road pricing system that charges drivers based on the distance they travel within the country \citep{Glavić2021}. The following section describes the cross-broader trend of car-traffic volume as shown in Figure  \ref{fig:annual-trend-cross}.
\begin{enumerate} 
    \item \textbf{At the German Border (observation station nearby Autobahn A1):} Figure 4 displays the monthly traffic volume percentages at the observation station located at the German border (Autobahn A1) for the years 2018 and 2020. Generally, the traffic volume during the morning rush hour is higher than that during the evening rush hour. By comparing the four plots, it can be seen that the highest morning-hour traffic volume was observed in September 2018. However, a sharp decrease in traffic volume was also noticed in August, during both morning and evening rush hours. During weekends, the highest traffic volume during morning hours was observed in May. In 2020, the weekday morning hour traffic volume reached its lowest point due to COVID-19-related restrictions. In 2018, the weekday evening-hour traffic volume increased, with the highest volume observed in September. From the end of October to November, a drop in the evening hour traffic volume was observed. In 2020, the weekday evening hour traffic volume reduced to nearly 50 percent, which may be due to COVID-19-related restrictions, except for the reduction observed in January.
    \item \textbf{At the French border (observation station near Autobahn A3):} The trend in traffic volume near the French border shows a similar pattern to that in Germany in terms of COVID-19 impact, with a drop at the end of March. However, the morning and evening traffic patterns are opposite, suggesting that more commuters are crossing the Luxembourg-French border during the evening rush hour. The detailed comparison of monthly and weekend traffic also varies greatly. For example, there was little variation in traffic volume between 2018 and 2020 in January. After the first wave of COVID-19 restrictions, the total traffic volume became similar. Although the highest volume was observed in December 2018, there was a sharp increase in evening traffic in October 2020. These results suggest that pandemic-related regulations may have a significant impact on traffic volume trends.
    \item \textbf{At the Belgium Border (observation station near Autobahn A7):} The annual trend chart displays the percentages of traffic volume for each month in 2018 and 2020. In 2018, traffic volumes were generally low during morning rush hours, with a peak observed in September. Weekday traffic volumes increased during morning rush hours, while traffic volumes on weekends decreased during evening rush hours. In contrast, in 2020, the traffic volume nearly doubled compared to 2018 for the entire year, with the lowest traffic observed from the end of March to April and the highest traffic observed from the end of September to October. This indicates a significant shift in traffic patterns between 2018 and 2020, likely due to the COVID-19 pandemic and related restrictions.
\end{enumerate} 
In 2018, traffic volume was very low traffic volume, whereas in 2020, the traffic volume was much higher. In contrast, the month of April 2018 had more traffic compared to April 2020. Towards the end of the year, the traffic volume in 2020 was higher than in 2018. During evening rush hours, the traffic volume was low, with a peak in October 2020 and September 2018. The traffic volume in the evening rush hours was significantly higher in 2020 compared to 2018. During weekend days, the traffic volume was very low in 2018, gradually increasing towards October 2020. By the end of the year, the traffic volume for both weekdays and weekends in 2020 was much higher than in 2018.

\subsubsection{Hourly Traffic (total) between Germany and Luxembourg}
The trend in traffic volume showed an exceptional pattern, prompting a physical data check that revealed numerous missing observations in the dataset. Consequently, further analysis will only focus on the observation station located at the German Border (crossing of Autobahn A1). This decision was made due to a particular interest in exploring the inter-country traffic volume between Germany and Luxembourg.

\begin{figure}[h!]
\centering
    \begin{subfigure}{0.48\textwidth}
        \centering
        \includegraphics[width=\textwidth]{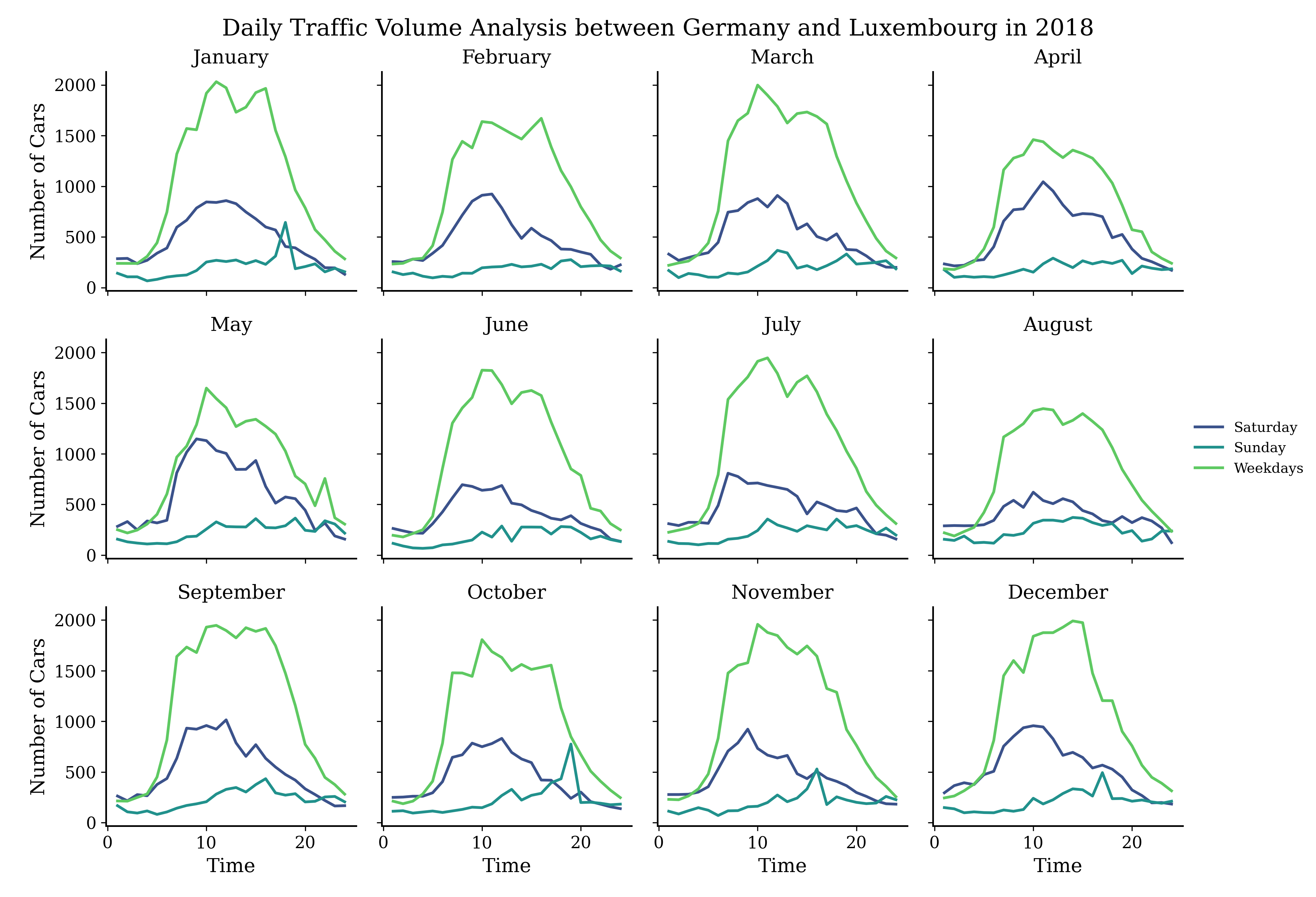}
        \caption{2018} 
        \label{fig:daily-trends-2018}
    \end{subfigure}
\centering
    \begin{subfigure}{0.48\textwidth}
        \centering
        \includegraphics[width=\textwidth]{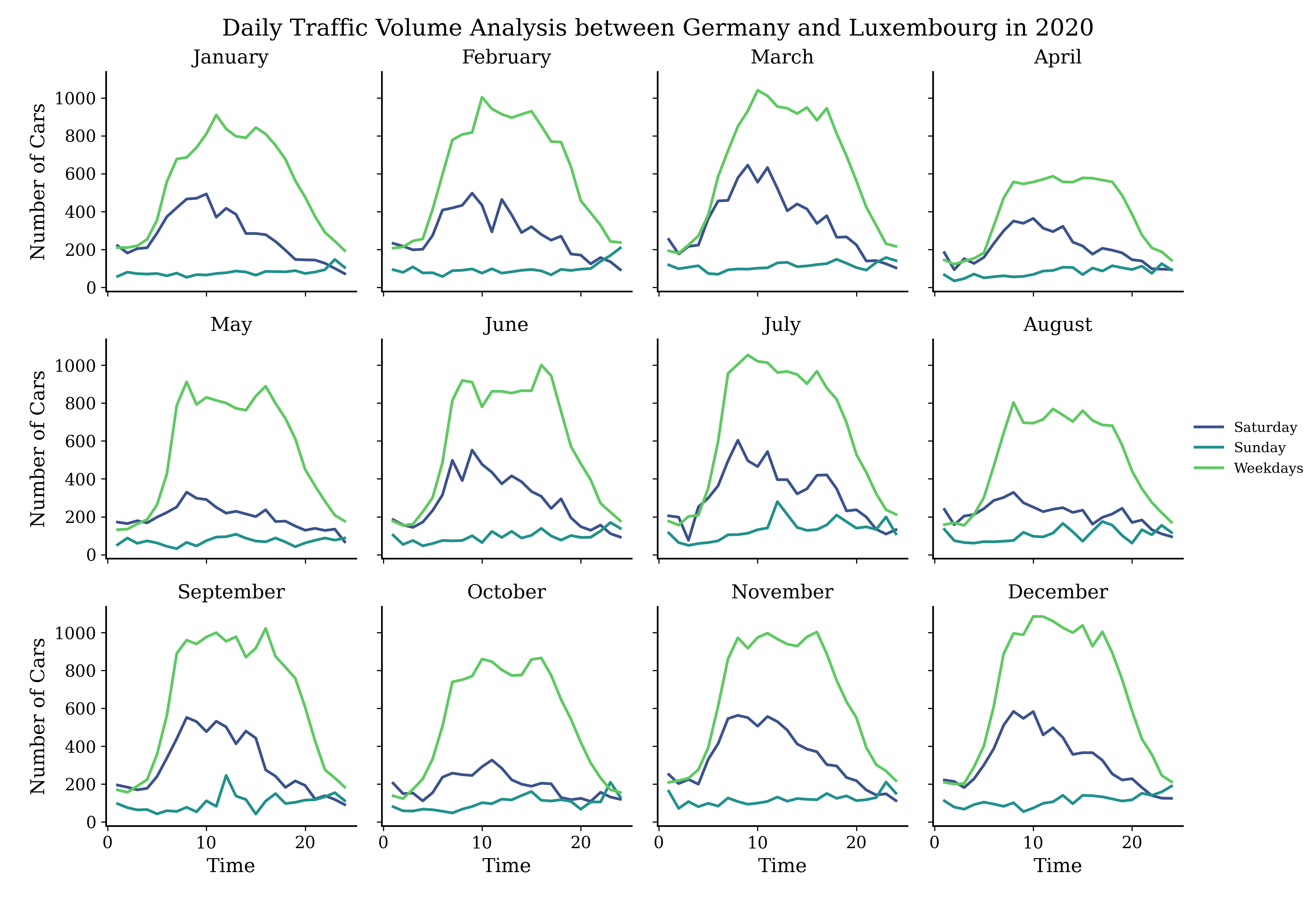}
        \caption{2020} 
        \label{fig:daily-trends-2020}
    \end{subfigure}
    \caption{Trend in daily (24 hours) traffic volume changes over the year in between Germany and Luxembourg in 2018 and 2020. Note: averaged count for weekdays.}
    \label{fig:daily-trends-2018-2020}
\end{figure}

Figure ~\ref{fig:daily-trends-2018-2020} illustrates the daily traffic volume between Germany and Luxembourg for each month in the years 2018 and 2020. In January, the traffic volume was higher in 2018 than in 2020. On the other hand, the traffic volume was higher in October and November of 2018 than in 2020. The peak traffic volume was observed on Sunday afternoons at 5:00 PM in January, October, November, and December, which may be linked to cross-border trips for recreational and tourism activities during the weekends. 

\subsubsection{Rush hours changes in Traffic volume by direction in percentage}

To analyse the morning and evening rush hour traffic, we considered the dataset's car volume between 7-10 AM and 4-7 PM, respectively. These time frames represent the typical morning and evening rush hours when traffic volume is typically at its highest. By focusing on these periods, we can gain a better understanding of the traffic patterns during peak commuting hours.

\begin{figure}[h!]
\centering
    \begin{subfigure}{0.49\textwidth}
        \centering
        \includegraphics[width=\textwidth]{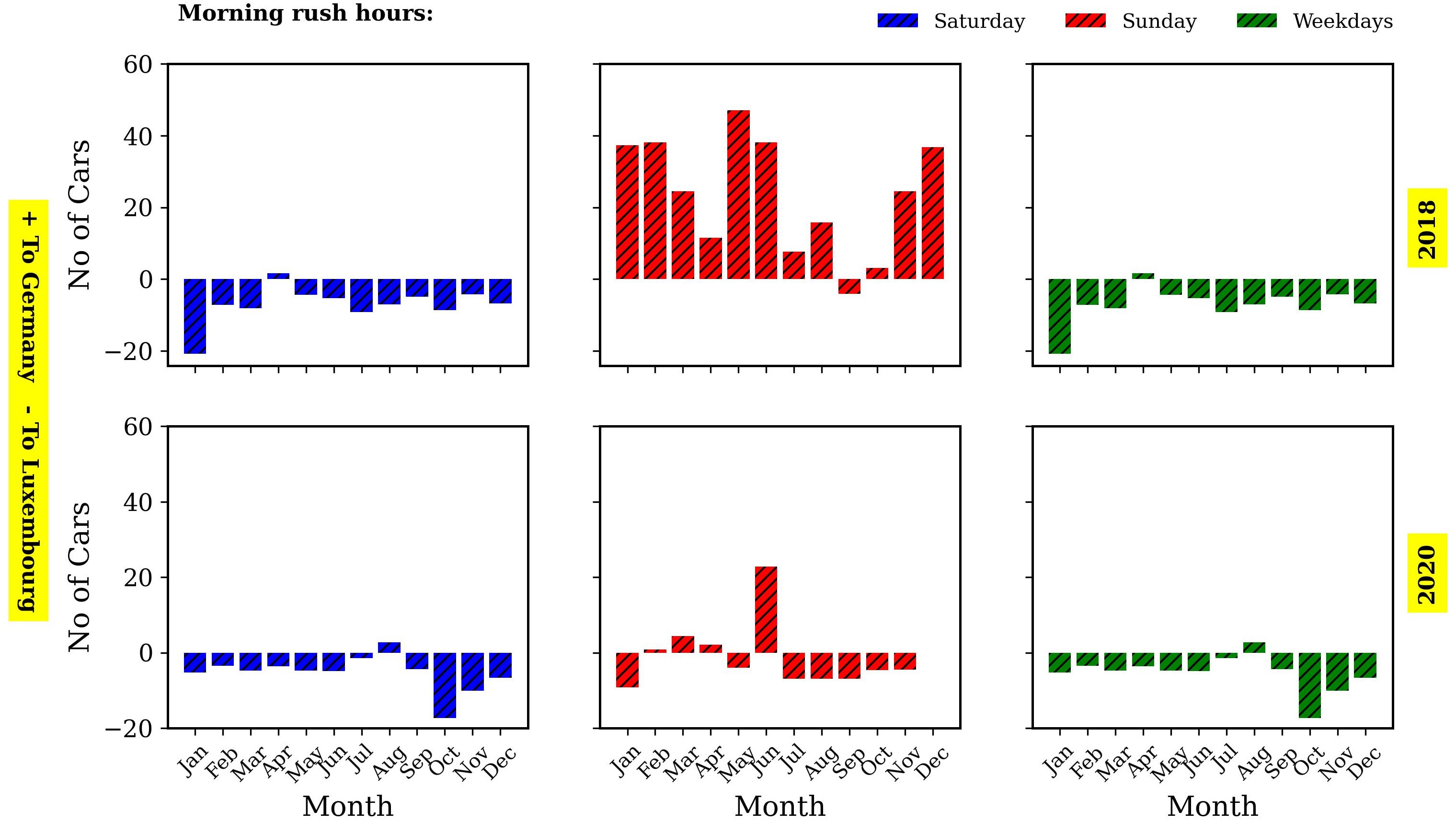}
        \caption{} 
        \label{fig:mor-rush-hours}
    \end{subfigure}
\centering
    \begin{subfigure}{0.49\textwidth}
        \centering
        \includegraphics[width=\textwidth]{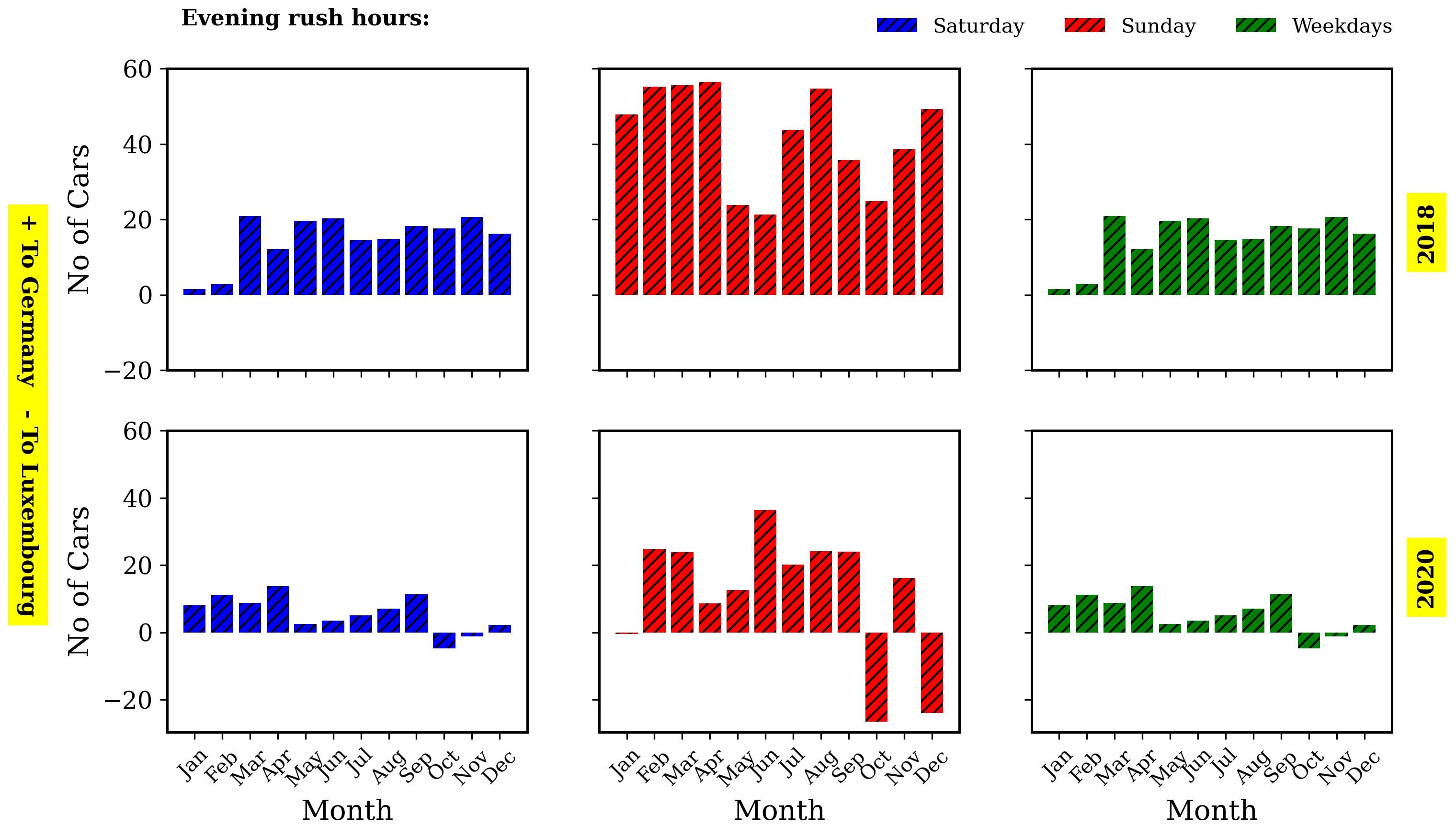}
        \caption{} 
        \label{fig:eve-rush-hours}
    \end{subfigure}
    \caption{Cross-sectional trend in traffic volume changes between Luxembourg to Germany during (a)morning and (b)evening rush hours.}
    \label{fig:mor-eve-rush-hours}
\end{figure}

Figure ~\ref{fig:mor-eve-rush-hours} displays a bar chart representing the percentages of traffic volume for each month between the years 2018 and 2020, along with the travel direction. The chart highlights a significant volume of car traffic on Sundays from Luxembourg to Germany, in both the morning and evening hours. In 2020, the evening rush hour on weekdays and Saturdays witnessed a slightly higher percentage of traffic volume from Germany to Luxembourg compared to the morning rush hour.
In 2018, the morning rush hour in January had the highest percentage of negative change, indicating that more people travelled from Germany to Luxembourg during that month. On Saturdays, the month of April showed the highest negative change. On Sundays, most months demonstrated a major change in the travel direction towards Germany. However, minimum changes were observed in traffic volume on weekdays.
In 2020, the car traffic volume had the highest percentage of negative change (i.e., travel direction to Luxembourg), with a major change in October and December. Saturday and weekdays showed the highest change, both positive and negative, compared to Sunday. Overall, the traffic volume in 2018 was higher than that in 2020, which could be strongly attributed to changes in travel behaviour due to COVID-19, among other factors.

\section{Discussion}
Our findings indicate a marked decrease in traffic volumes during the COVID-19 pandemic, which aligns with global observations of reduced mobility due to lockdowns and remote work policies. This reduction in traffic not only highlights the adaptability of travel behaviours in response to external shocks but also provides a unique opportunity to study the potential long-term implications of such behavioural shifts. This study demonstrates that the national open traffic data products have an impact beyond the national broader (e.g. Luxembourg only provides open traffic data, but Germany does not provide open traffic count data) and have adoption potential to address cross-border issues. 

The study introduced an analytical framework for sensing road traffic volumes at cross-border regions and incorporates a multi-dimensional approach (see section 3). Each dimension of the framework addresses specific aspects of data collection, processing, and analysis. The data quality dimension emphasises the importance of visualising and statistically analysing traffic data to understand its quantity and distribution over different periods and locations. High-quality data is essential for deriving meaningful insights and making informed decisions \citep{Chaloli2019, Taboada2020}. The data harmonisation dimension is crucial for dealing with issues such as missing entries, unequal entry counts, and the aggregation of data at various temporal resolutions. By creating secondary data and performing sub-setting, this dimension ensures that the dataset is robust and suitable for detailed analysis. The distributional dimension focuses on identifying emerging trends, hot spots, and spatial intensity patterns in traffic data. By analysing these patterns at both local and national scales, we can better understand traffic flow dynamics and identify critical areas requiring attention.  The directional cross-border dimension explores the complexity of cross-border traffic flows, particularly their temporal variations and directional patterns. By concentrating on major entry points near international borders during peak hours, this dimension aims to capture the daily commuter traffic for diverse trip purposes. 

The study methodology is particularly suited for car-based travel pattern analysis in cross-border regions using national open data. The practical case study demonstrated how to leverage multiple data sources for investigating travel behaviours and patterns even beyond the national broader. The data quality and harmonisation dimensions ensure that the dataset is reliable and well-structured, enabling accurate trend analysis and pattern detection. Moreover, the distributional dimension allows for the identification of hot spots and critical traffic clusters, which are essential for urban planning and traffic management \citep{Hullman2021, Hamza2023}. The directional cross-border dimension provides insights into the unique challenges of managing traffic flows at international borders, highlighting the importance of targeted data collection and analysis in these areas. Beyond travel pattern analysis, the analytical framework has broad applicability for various types of data analysis \citep{Makaba2020}. Its multi-dimensional approach can be adapted to different domains where large, complex datasets are involved. The emphasis on data quality, harmonisation, and distributional analysis ensures that the framework can handle diverse data types and provide actionable insights \citep{Hullman2021}. For instance, in environmental monitoring, similar dimensions can be used to analyse pollution data, identify hot spots of high pollutant concentrations, and understand temporal and spatial variations. In healthcare, the framework can be adapted to study disease outbreak patterns, patient demographics, and healthcare resource utilization \citep{deHaas2020, Eisenmann2021}. The study approach is transferable to other investigations that require detailed data analysis with traffic count data and interpretation. By integrating various data sources and focusing on key analytical dimensions, the framework can help researchers and practitioners gain a deeper understanding of complex phenomena and make informed decisions.
This study of car-based human mobility behaviour mostly focused on quantified observations and fact analysis with limited statistically significant testing. The actionable insights can be formulated by including the correlation analysis of mobility changes with policy interventions or socio-economic factors. It could be extended to further study in adoption of triangulation approaches for robust decision-making. Future research should extend this analysis to include traffic between Belgium and France to offer a more comprehensive view of regional mobility trends. Additionally, detailed interpolation and downscaling approaches are necessary to pinpoint pollution hotspots and calculate spatial influence areas, thereby enhancing our understanding of the environmental impact of car-based travel.
The correlation of traffic data with air pollution levels could provide critical insights into the contribution of cross-border mobility to air quality issues. This is particularly relevant in the context of the net-zero carbon transition, where reducing emissions from transportation is a key objective. Interdisciplinary collaborations will be essential to advance this research. By integrating expertise from transportation engineering, environmental science, and data analytics, we can better understand mobility patterns and their broader implications. This study lays the groundwork for such interdisciplinary efforts, offering valuable insights into the evolving dynamics of car-based human mobility in the pandemic era.

\section{Conclusion}
This study presents an explorative approach in the adoption of open traffic data to analyse spatio-temporal changes in car-based human mobility using high-frequency data from IoT sensor networks. An analytical framework has been proposed and demonstrated in an application with a real-world case study at the cross-border area of Luxembourg-Germany. A reproducible pipeline used for data processing, missing data handling and computing the traffic flow variables. Our study reports the explorative trends in individual car-based daily (24h) travel behaviour before (2018) and during the COVID-19 pandemic (2020) between Luxembourg and neighbouring countries. Although this study explores the inter-country traffic volume between Germany and Luxembourg in detail, further investigation could include a similar approach for analysing traffic volume between Belgium and France. The traffic crossing multiple routes might be an interesting statistic for understanding temporal road users, which can also be studied in combination with other information using this dataset.

Since February 2020, Luxembourg has introduced a new road pricing system called "MyMove," which charges drivers based on the distance they travel in the country. There are certain exemptions and reductions available for commuters who frequently travel to and from Luxembourg for work purposes. Our replicable workflow and analytics may show some impact on such a policy that aims to reduce traffic congestion and promote sustainable mobility. Such topics need to be investigated comprehensively through interdisciplinary collaborations.

In relevance to the net-zero carbon transition \cite{seto2021low}, future studies should shed light on interpolation and downscaling approaches at the comprehensive road network level for identifying pollution hot spots and calculating the spatial influence area. Correlation analysis can also be conducted, for example, using traffic count variables to investigate the spatial impact of air pollution due to the increase and decrease in car volume. In addition, future work could also implement a more advanced analytical framework integrating Granger causality modeling to predict air pollution with greater precision \citep{Sinha2023}. The findings can be useful in understanding how much cross-border travellers are responsible for air pollution related to car-based travel.

\section {Declarations}

\begin{itemize}
    \item \textbf{Funding:} The authors did not receive support from any organization for the submitted work.
    \item \textbf{Clinical trial number:} not applicable.
    \item \textbf{Ethics approval and consent to participate}: Not applicable. 
    \item \textbf{Consent for publication}: Not applicable.
    \item \textbf{Author Contribution declaration:} SKS: Conception, Data analysis, Formal analysis, Writing, supervision, project lead; JI:  Data processing, Coding, Writing; HL: Validation, editing; HO: Data collection, editing.
    \item \textbf{Competing interests:} The authors declare no competing interests.
    \item \textbf{Data availability:} Get open and free access to  data, code, visualization and documentation in: \href{https://doi.org/10.5281/zenodo.13383896}{DOI link}
    
\end{itemize}


\bibliography{ref_human_mobility}

\begin{appendices}
\section{Supplementary Images}

\begin{figure}[htb]
\centering
    \begin{subfigure}{0.48\textwidth}
        \centering
        \includegraphics[width=\textwidth]{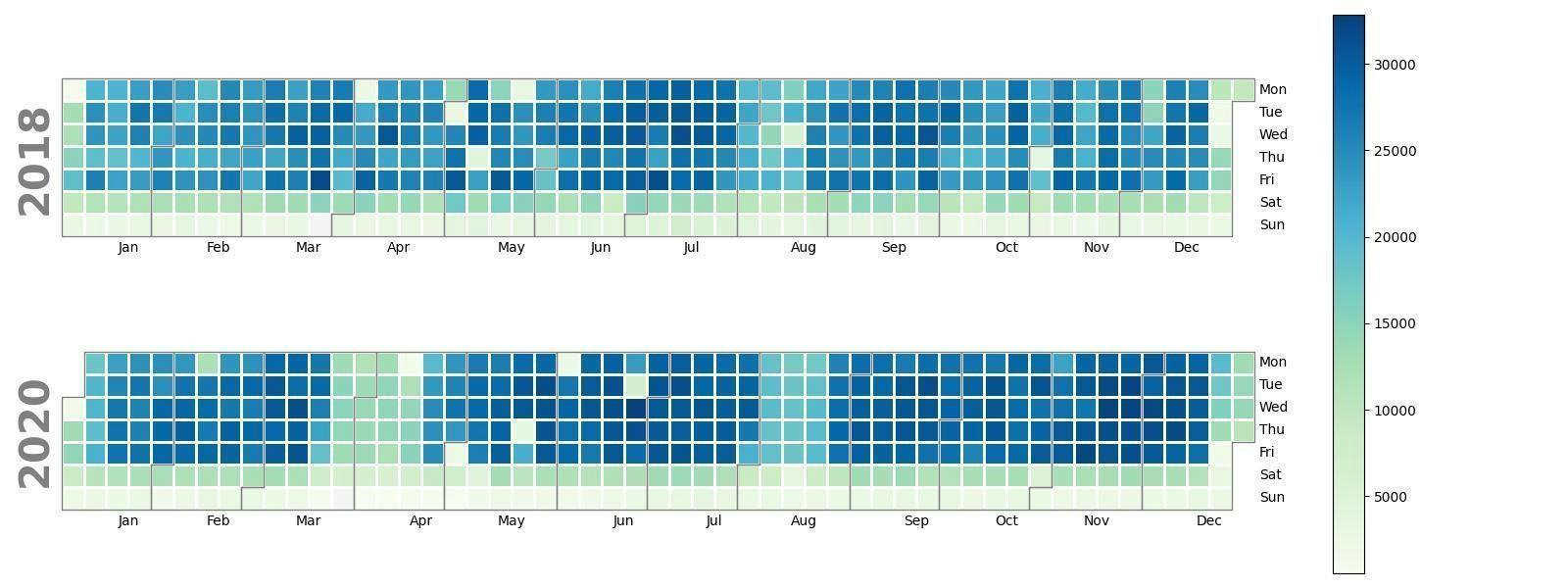}
        \caption{Morning 7:00 AM 10:00 AM} 
        \label{fig:cal-morning}
    \end{subfigure}
\centering
    \begin{subfigure}{0.48\textwidth}
        \centering
        \includegraphics[width=\textwidth]{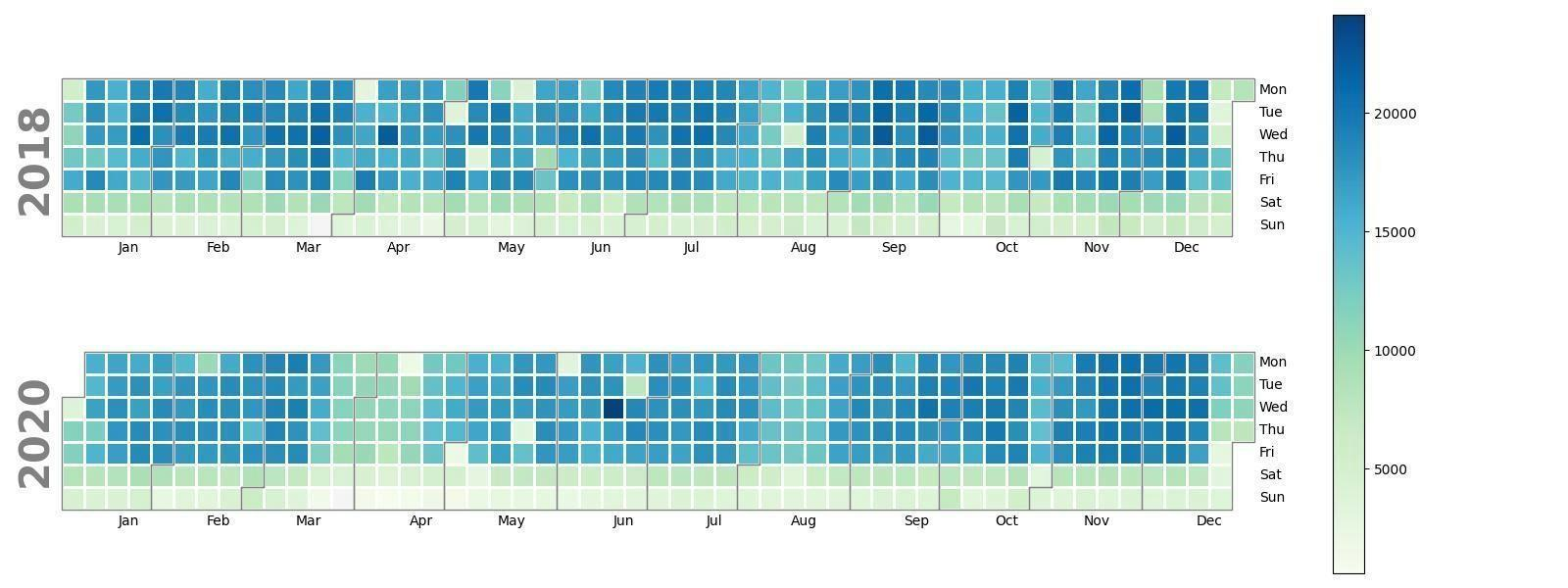}
        \caption{Evening 16:00 PM to 19:00 PM} 
        \label{fig:cal-morning}
    \end{subfigure}
    \caption{Calendar chart of traffic volume in 2018 and 2020}
    \label{fig:cal-morning-evening}
\end{figure}

\begin{figure}[htb]
\centering
    \begin{subfigure}{0.48\textwidth}
        \centering
        \includegraphics[width=\textwidth]{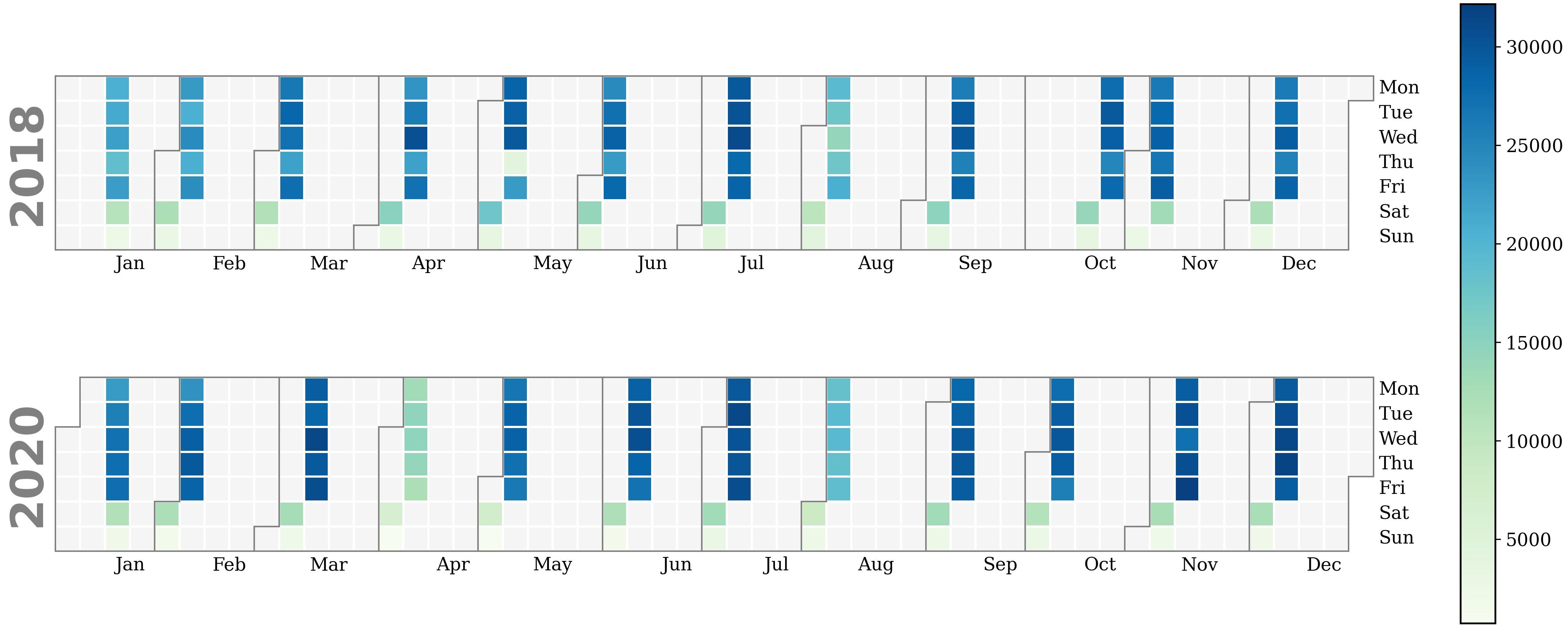}
        \caption{Morning 7:00 AM 10:00 AM} 
        \label{fig:cal-morning-sel}
    \end{subfigure}
\centering
    \begin{subfigure}{0.48\textwidth}
        \centering
        \includegraphics[width=\textwidth]{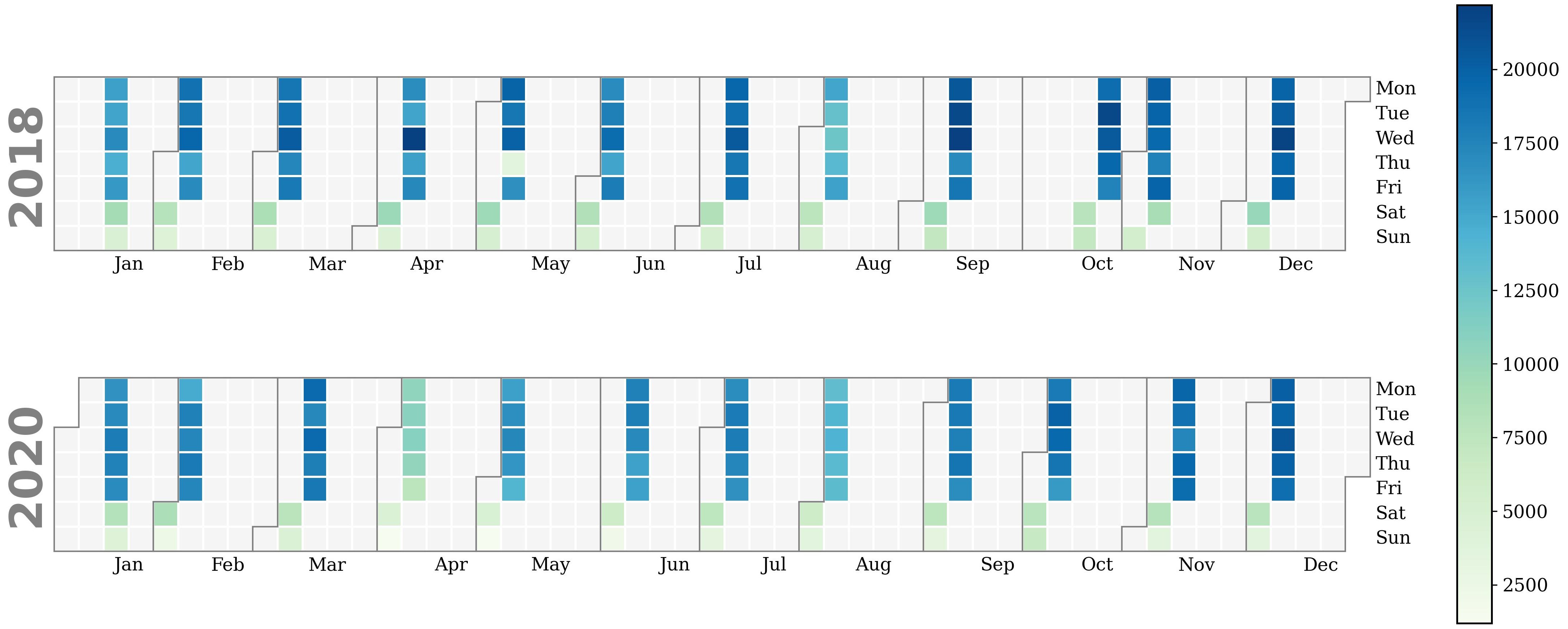}
        \caption{Evening 16:00 PM to 19:00 PM} 
        \label{fig:cal-morning-sel}
    \end{subfigure}
    \caption{Selected weeks per month that provide best completeness in data}
    \label{fig:cal-morning-evening-sel}
\end{figure}

\begin{figure}
    \centering
    \includegraphics[width=0.85\linewidth]{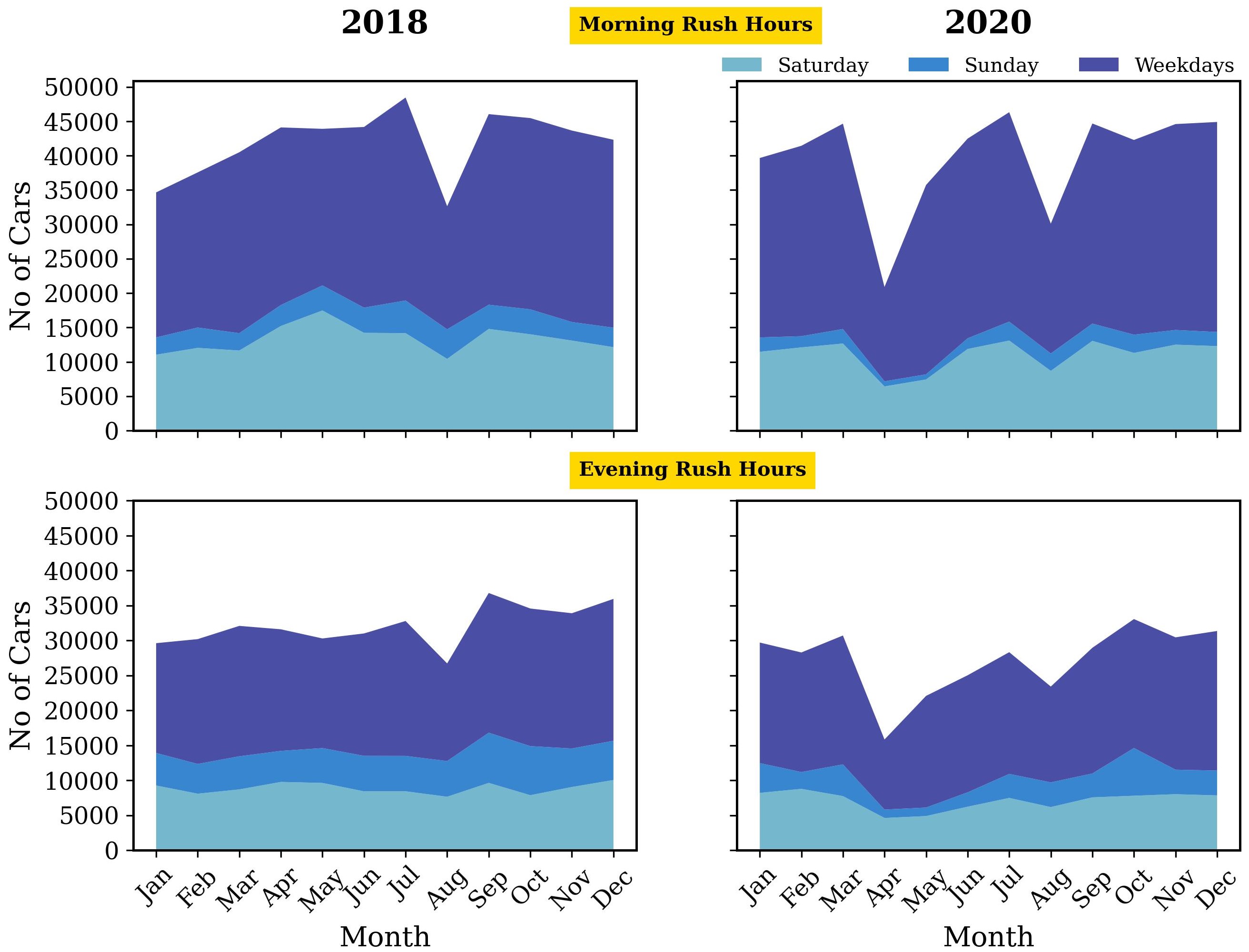}
    \caption{Comparative trend in total car-traffic volume in between 2018 and 2020}
    \label{fig:traffic-volume}
\end{figure}

\begin{figure}
    \centering
    \includegraphics[width=0.9\linewidth]{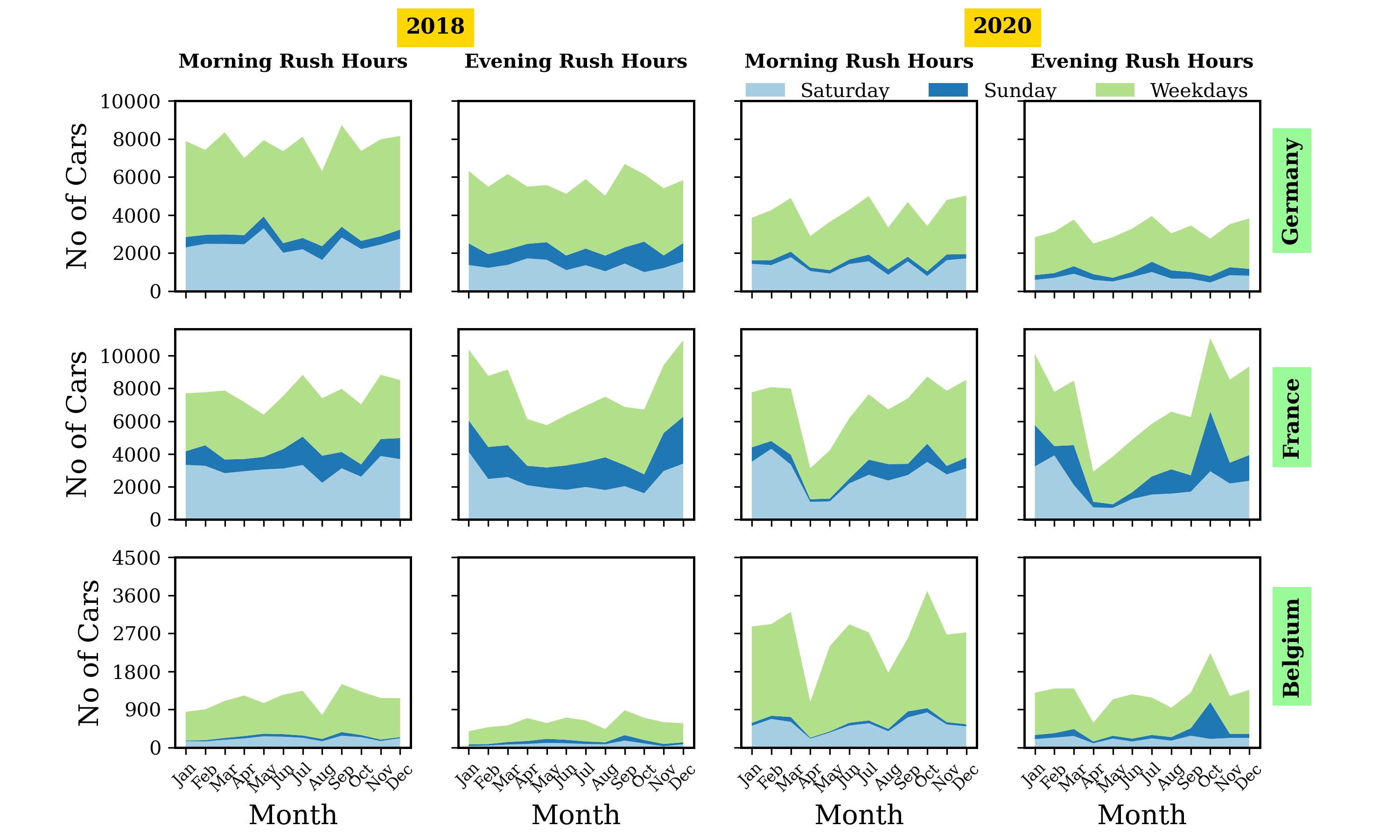}
    \caption{Annual trend in total traffic volume at the cross-border observation stations}
    \label{fig:annual-trend-cross}
\end{figure}

\end{appendices}






\end{document}